\documentclass[10pt]{iopart} 

\usepackage[latin1]{inputenc}
\usepackage{graphicx}
\usepackage[colorlinks,citecolor=blue,urlcolor=blue,linkcolor=blue]{hyperref}
\usepackage{siunitx}

\expandafter\let\csname equation*\endcsname\relax
\expandafter\let\csname endequation*\endcsname\relax

\usepackage{amsmath}
\usepackage{braket}

\usepackage{iopams}
\usepackage{widetable}
\usepackage{bm}
\usepackage{cite}
\usepackage{dsfont}
\usepackage{booktabs}
\usepackage{supertabular}

\newcommand{\op}[1]{\hat{\textbf{#1}}}
\newcommand{\opd}[1]{\hat{\textbf{#1}}^{\dagger}}

\begin{document}

\title[]{Broadband indistinguishability from bright parametric downconversion in a semiconductor waveguide}

\author{T. G\"unthner$^1$, B. Pressl$^1$, K. Laiho$^1$, J. Ge\ss ler$^2$, S. H\"ofling$^{2,3}$, M. Kamp$^2$, C. Schneider$^2$ and G. Weihs$^{1, 4}$}

\address{$^1$Institut f\"ur Experimentalphysik, Universit\"at Innsbruck, Technikerstra\ss e 25, 6020 Innsbruck, Austria}

\address{$^2$Technische Physik, Universit\"at W\"urzburg, Am Hubland,  97074 W\"urzburg, Germany}

\address{$^3$School of Physics $\&$ Astronomy, University of St Andrews, St Andrews, KY16 9SS, United~Kingdom}

\address{$^4$Institute for Quantum Computing, University of Waterloo, 200 University Avenue W, Waterloo, ON, N2L 3G1, Canada}

\eads{\mailto{thomas.guenthner@uibk.ac.at} and \mailto{kaisa.laiho@uibk.ac.at}}

\date{\today}

\begin{abstract}
Parametric downconversion (PDC) in semiconductor Bragg-reflection waveguides (BRW) is routinely exploited for photon-pair generation in the telecommunication range. Contrary to many conventional PDC sources, BRWs offer possibilities to create spectrally broadband but nevertheless indistinguishable photon pairs in orthogonal polarizations that simultaneously incorporate high frequency entanglement. We  explore the characteristics of co-propagating twin beams created in a type-II ridge BRW.  Our PDC source is bright and efficient, which serves as a benchmark of its performance and justifies its exploitation for further use in quantum photonics. We then examine the coalescence of the twin beams and investigate the effect of their inevitable multi-photon contributions on the observed photon bunching.  Our results show that BRWs have a great potential for producing broadband indistinguishable photon pairs as well as multi-photon states.
\end{abstract}

\pacs{42.65.Lm, 42.50.-p, 42.50.Dv, 42.65.Wi }

\submitto{\JOPT}


\ioptwocol

\section{Introduction}
\label{sec:intro}
Versatile quantum light sources are needed for a variety of quantum communication tasks, and thus 
we would like to develop them for the telecommunication wavelengths. Perhaps the best scrutinized process is parametric downconversion (PDC), which produces photons in pairs, usually denoted as signal and idler. Waveguide realizations have turned PDC sources into easy-to-handle and small-scale tools that moreover provide higher brightness than their bulk counterparts \cite{Tanzilli2002, Fiorentino2007}. Waveguided sources further provide better integrability and quantum integrated networks have been built both on semiconductor as well as ferro-  and dielectric platforms \cite{Krapick2013, Silverstone2013, Wang2014, Krapick2014}.

Recently, we and others demonstrated PDC in waveguides being composed of layers of semiconductor materials that are historically named Bragg-reflection waveguides (BRWs) \cite{Yeh1976, Lanco2006, Horn2012}.
In comparison to ferroelectric non-linear optical waveguides the semiconductor structures benefit from higher nonlinearity and better integrability \cite{Horn2012}. 
By embedding the pump laser and the photon-pair production on the same chip the PDC emission can even be electrically self-pumped \cite{Bijlani2013, Boitier2014}. 
BRWs with high signal-idler correlations are suitable for various quantum optical tasks \cite{Horn2012, Sarrafi2013}. Previous experimental studies include the investigation of the photon-pair indistinguishability \cite{Caillet2010} and preparation of polarization entangled states both with co- and counter-propagating signal and idler schemes \cite{Orieux2013, Horn2013, Valles2013}. 
Furthermore, BRWs  offer a lot of flexibility for source design, which aims at engineering of quantum states with desired properties for specific applications \cite{Thyagarajan2008, Abolghasem2009, Svozilik2011,Svozilik2012, Kang2012, Eckstein2014}.

However, in order to successfully compete with conventional PDC sources, BRWs have to be able to produce twin beams, in other words signal and idler obeying a strict photon-number correlation, in a bright and efficient manner with a low number of spurious counts \cite{Lanco2006, Horn2012}.
Still today, their drawbacks are the incompatibility of the utilized spatial modes with standard single-mode fiber optics, a rather high facet reflectivity because of the large refractive index difference with air, and a high numerical aperture (NA)  due to the strong confinement of the spatial modes. On top of this, the optical losses both at the pump and the downconverted wavelengths are significant \cite{Bijlani2009, Horn2012}, which limits the useful length of the structures.

The phasematching required for the PDC process can be achieved in semiconductor waveguides by spatial mode matching \cite{Anderson1971, Banaszek2001, Helmy2011} eliminating the need for quasi-phasematching, which is typical for conventional sources. 
All in all, the characteristics of signal and idler in their different degrees of freedom are determined by the PDC process parameters such as the strength of the nonlinearity, pump envelope and the dispersion of the interacting modes in the used geometry. 
The resulting joint spectrum of signal and idler to a large extent dictates for which quantum optics applications the photonic source in question is suitable \cite{Grice1997, U'Ren2005, Zhukovsky2012}. 
The state-of-the-art BRW sources provide high entanglement in the spectral degree of freedom \cite{Zhukovsky2012b, Avenhaus2009, Brida2009}. Simultaneously, they  also offer broadband spectral indistinguishability for signal and idler  that are created in orthogonal polarizations. The former is desired in applications requiring multimode PDC characteristics---or higher dimensional states \cite{Bernhard2013, Wakui2014}, whereas the latter is a building block for many quantum optical networks that base on photon bunching \cite{Hong1987}. 
The PDC process parameters also govern the photon statistics of the twin beams, inevitably resulting in higher photon-number contributions, which have to be controlled \cite{Avenhaus2008}.
 
Here, we investigate the characteristics of spectrally broadband type-II PDC emitted by a BRW in a single-pass configuration.  First, we determine the Klyshko efficiency of our BRW source. Thereafter, we utilize correlation functions between signal and idler in order to investigate the mean photon number in them. We further test the broadband indistinguishability in a two-photon coalescence experiment. 
For this purpose, we manipulate the spectral bandwidth of signal and idler by broadband filtering, and via their bunching at different gains we determine the indistinguishability governed by the spectral overlap. Our results show that BRWs are bright and efficient photon sources.

\begin{figure*}[!htb]
\centering
\includegraphics[width=\linewidth]{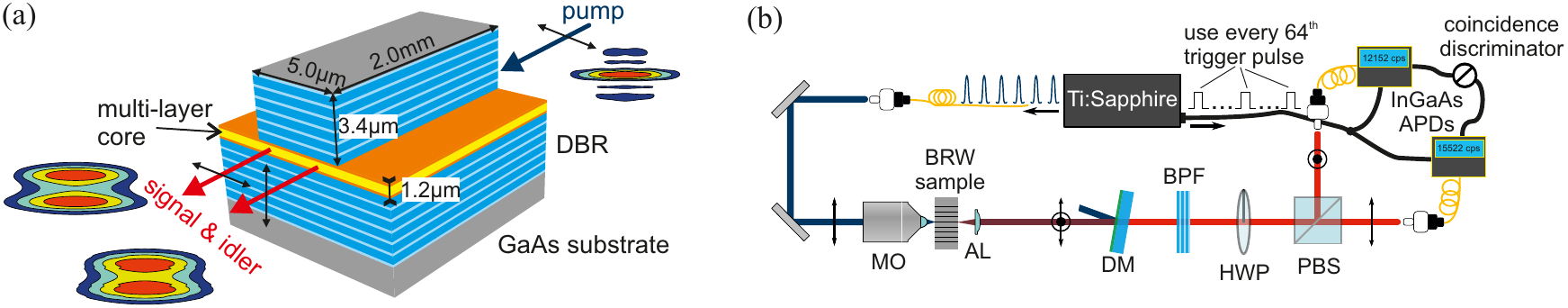} 
\caption{(Color online) (a) Investigated BRW sample with a core thickness of \SI{1.2}{\micro\meter}, a length of \SI{2.0}{\milli\meter} and a ridge width and height of \SI{5.0}{\micro\meter} and \SI{3.4}{\micro\meter}, respectively. The depicted amplitude distributions of pump, signal and idler mode were inferred with a commercial-grade simulator eigenmode solver and propagator \cite{Mode06}. 
(b) Experimental setup for investigating the  source efficiency and brightness as well as the signal and idler coalescence. For abbreviations and more details see text. 
}
\label{fig:exp}
\end{figure*}
\section{Sample design and experiment}
\label{sec:sample}
Our BRW sample depicted schematically in \fref{fig:exp}(a) is grown on a GaAs substrate by molecular beam epitaxy. The sample having the same structural design as in \cite{Horn2012, Horn2013} is made of Al${}_{x}$Ga${}_{1-x}$As ($0<x<1$) compounds because of their inherent high optical second order nonlinearity and the sophisticated fabrication techniques available. Due to its zincblende structure, Al${}_{x}$Ga${}_{1-x}$As has no birefringence and, therefore, in order to  achieve phasematching  the sample is designed to support different spatial modes that are the Bragg mode for the pump  and the total internal reflection (TIR) modes for the twin beams  \cite{West2006}. The distributed Bragg reflectors (DBRs) embed a multi-layer core that guarantees good spatial overlap of the pump, signal and idler mode triplets required for an efficient PDC process \cite{Mizrahi2004, Christ2009, Abolghasem2010}. Finally, the ridge structure is fabricated by electron beam lithography followed by plasma etching. This ensures mode confinement in two dimensions---in the vertical direction by the DBRs and in the horizontal direction by the ridge structure.

In our experiment as shown in \fref{fig:exp}(b) we employ a picosecond pulsed Ti:Sapphire laser ({\SI{76.2}{\mega\hertz} repetition rate, \SI{772}{\nano\meter} central wavelength) as a pump for the PDC process. After power, polarization and spatial mode control, we focus the pump on to the front facet of our BRW with a 100x microscope objective (MO), which allows reasonably efficient coupling into the Bragg mode. At its output facet a high NA aspheric lens (AL) collects the PDC emission from our BRW, for which we numerically determined the NAs of approximately $0.2$ and $0.5$ in the horizontal and vertical direction, respectively \cite{Mode06}.
After removing the residual pump beam with a dichroic mirror (DM), we use a band-pass filter (BPF) with either a \SI{12}{\nano\meter} or \SI{40}{\nano\meter} bandwidth to spectrally limit the twin beams.
With an additional half-wave plate (HWP) we change the polarization direction of signal and idler, as necessary for the coalescence experiment.
After separating the orthogonal polarizations using a polarizing beam splitter (PBS), we launch each beam to a single-mode fiber for detection with two commercial time-gated InGaAs avalanche photo-diodes (APDs). By utilizing a narrowband telecom laser for adjustment, we estimate that the TIR modes can be coupled with approximately 50\% efficiency into the used fibers. Additionally, the variable photodetection probability of our APDs is set either to \SI{20}{\percent} or \SI{25}{\percent} resulting in dark count rates of about \SI{70}{\second^{-1}} and \SI{200}{\second^{-1}}, correspondingly.  Finally, we employ a time to digital converter to discriminate the single and coincidence counts. Due to technical limitations, only every 64$^{\text{th}}$ laser pulse gates our InGaAs APDs corresponding to a rate $R$ of about \SI{1.19}{\mega\hertz}.

\section{Source efficiency and brightness}
\label{sec:results1}

With conventional PDC sources efficient single-mode fiber couplings  and  large mean photon numbers can be achieved at telecommunication wavelengths  \cite{Zhong2009, Harder2013, Eckstein2011, Wakui2014}. 
Therefore, we start by investigating the performance  of our BRW source by determining its Klyshko efficiency \cite{Klyshko1977} and thereafter evaluate the mean photon number in the individual twin beams.
For this purpose we use the configuration shown in \fref{fig:exp}(b), in which the PDC emission is filtered to a  spectral width of  \SI{40}{\nano\meter} to suppress background light from the waveguide before the signal and idler beams are deterministically separated at PBS.
\begin{figure}[!tb]
\centering
\includegraphics[scale=1.1]{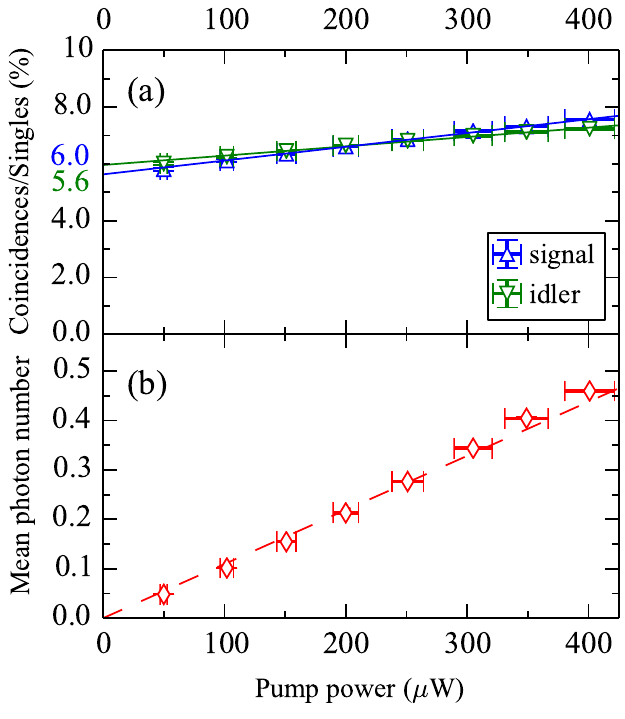}
\caption{(Color online) (a) The ratio of coincidences to singles in both signal and idler and (b) mean photon number $\langle n \rangle$ in one of the generated twin beams with respect to the pump power. The vertical errorbars are smaller than the used symbols.}
\label{fig:res1}
\end{figure}

To eliminate the effect of the accidental coincidences produced by the higher photon-number contributions created in PDC, we measure the process efficiency with respect to the pump power. In the region of weak pump powers we can extract the Klyshko efficiency, which is defined  only for perfectly photon-number correlated photon-pairs, as
$\eta_{{s,i}} = C/S_{i,s}$
with $C$ being the coincidence rate and $S_{s,i}$ the single count rates of signal and idler, respectively. Our results in \fref{fig:res1}(a) show the ratio of  coincidence counts to single counts for both signal and idler detected with a chosen photon detection probability of \SI{20}{\percent} at our APDs. Thence, we extrapolate Klyshko efficiencies of \SI{6.0(1)}{\percent} and \SI{5.6(2)}{\percent} for signal and  idler, respectively. This indicates that despite the high NA and the complex spatial mode structure, the PDC emission can be fairly efficiently collected with standard single-mode optics even when compared with the performance of the conventional sources \cite{Harder2013, Eckstein2011}.

By extracting the signal and idler cross-correlation $C/A$, in which $A = S_{i} S_{s} / R$ corresponds to the accidental count rate, we can further estimate the mean photon number $\braket{n}$ created in one of the twin beams in a loss-independent manner via $C/A \approx 1/\braket{n} + 1$  \cite{Christ2011}.
Since our BRW is a highly multimodal PDC source (see~\ref{app1}), this estimate gives the mean photon number in good approximation, being in the worst case the lower bound.
In \fref{fig:res1}(b) we illustrate the obtained mean photon number growing linearly with respect to the increasing pump power as expected for weakly excited PDC. Our results show that mean photon numbers up to $0.5$ can be achieved with moderate pump powers. This further indicates that  the PDC multi-photon contributions have to be taken into account especially in the photon coalescence experiment we investigate next.


\section{Coalescence of signal and idler}
\label{sec:results2}
For observing the coalescence of signal and idler photons we follow the experiment in \cite{Avenhaus2009} and investigate photon bunching by varying their distinguishability in the polarization degree of freedom. For this purpose, we detect the coincidences between signal and idler while rotating the HWP in \fref{fig:exp}(b). In case the HWP axes are oriented parallel to the cross-polarized signal and idler, they are separated deterministically at the PBS. 
In any other case, signal and idler bunch together if they are indistinguishable in all degrees of freedom---not only in polarization but also spatially and spectrally. 

We record the coincidence counts with respect to the HWP angle at several pump powers. Additionally, we change the photon detection probability of our APDs from \SI{20}{\percent} to \SI{25}{\percent} to increase  the count rates.
\Fref{fig:res2}(a) and (b) show our results with a \SI{12}{\nano\meter} band-pass filter (BPF) for a low and a high pump power value, respectively. From this, we can directly conclude that the higher the pump power the lower is the visibility of the measured fringes given by $\mathcal{V}=(C_{\mathrm{max.}} - C_{\mathrm{min.}}) / (C_{\mathrm{max.}} + C_{\mathrm{min.}})$. We nevertheless achieved a maximum visibility of \SI{0.83(1)}{}, which lies significantly above the classical limit of $1/3$ for a completely distinguishable photon pair with no higher-order contributions \cite{Avenhaus2009} and delivers a measure of the signal and idler indistinguishability.
\begin{figure}[!htb]
\centering
\includegraphics[scale=1.1]{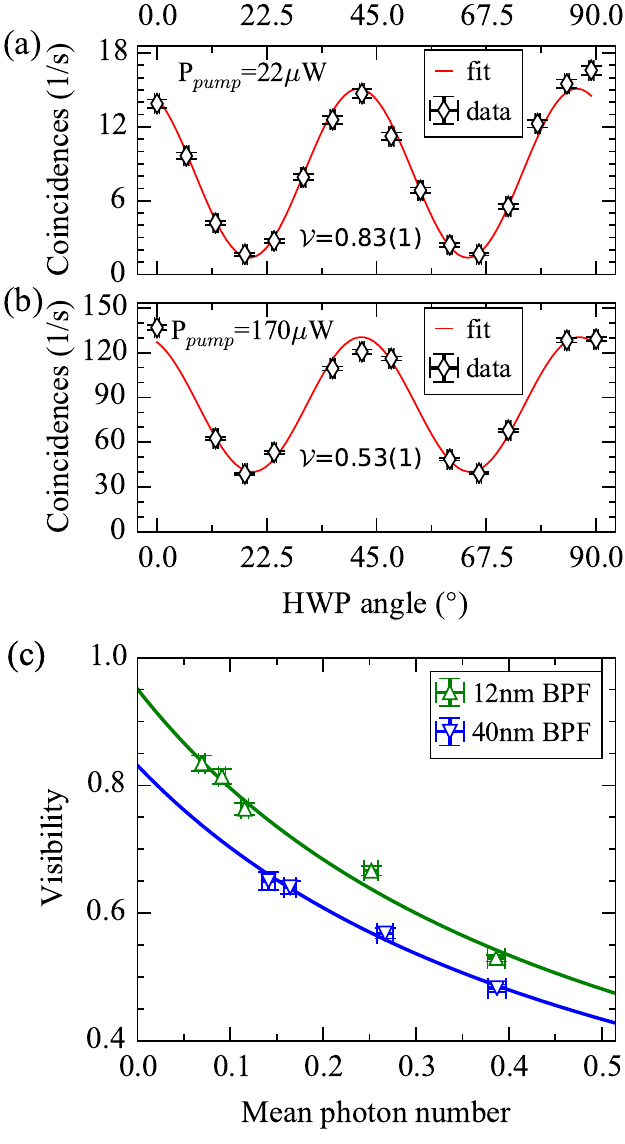}
\caption{(Color online) Measured coincidence counts as function of the HWP angle for pump powers of (a) \SI{22}{\micro\watt} and (b) \SI{170}{\micro\watt} for a \SI{12}{\nano\meter} filter bandwidth. (c) The extracted visibilities decrease with increasing mean photon number.}
\label{fig:res2}
\end{figure}

In order to understand the effect of the multi-photon contributions to the signal and idler coalescence, we further examine the visibility in terms of the mean photon number, which is also directly provided  by the measured data, when signal and idler are deterministically split at HWP. 
In \fref{fig:res2}(c) we depict the measured visibilities with respect to the estimated mean photon numbers for both \SI{12}{\nano\meter} and \SI{40}{\nano\meter} BPFs.
In both cases  the visibility clearly decreases with increasing mean photon number. 
However, not only the increasing multi-photon contributions but also the spectral mismatch between signal and idler affect the visibility in the coalescence experiment. From the measured visibilities we can infer the spectral overlap of the downconverted photon pairs  via (see \ref{app2})
\begin{equation}
\mathcal{V} \approx \frac{1 + \mathcal{O} }{ 3 - \mathcal{O} + 4 \langle n \rangle }.
\label{eq:visfit}
\end{equation}
By fitting our results in \fref{fig:res2}(c)  against \eref{eq:visfit} we retrieve for the spectral overlap  with \SI{12}{\nano\meter} and \SI{40}{\nano\meter} filter bandwidths the values of \SI{95.0(6)}{\percent} and \SI{81.6(9)}{\percent}, respectively. The large difference in the spectral overlaps with the two investigated filters are caused by the fact that signal and idler wavepackets are temporally shifted with respect to each other due to their slightly different group velocities. Theoretically, the \emph{joint spectral distribution} of signal and idler governs the spectral overlap (see \ref{app1}). Our results are in good accordance with numerical simulations, which predict spectral overlaps of about \SI{98}{\percent} and \SI{83}{\percent}, respectively for the two filters.

\section{Conclusion}
\label{sec:conclusions}

Integratable and easy-to-handle sources of parametric downconversion are highly desired in many quantum optics applications. Bragg-reflection waveguides based on semiconductor compounds provide a platform that can meet these demands. 
Our BRW sample shows a good performance, and we can reach Klyshko efficiencies up to a few percent with avalanche photodetection, regardless of the non-standard mode profile of the signal and idler beams. Moreover, our source provides a high brightness and is capable of producing higher photon numbers as is desired for multiphoton production.  We further examined the coalescence between the twin beams filtered to a few tens of nanometers bandwidth in order to assess their indistinguishability. The visibility of the measured fringes is diminished by the multi-photon contributions of signal and idler, but  we can nevertheless extract a high degree of indistinguishability, which is quantified by their spectral overlap. We extended our model to take into account both these process parameters and showed that our results are in good agreement with numerical simulations. Thus, being characteristic for BRWs, our source provides signal and idler in orthogonal polarizations that are over a broad spectral band highly indistinguishable in frequency. We believe our work gives a detailed insight of the PDC process in our BRW both in the spectral and photon-number degrees of freedom. This will become important when optimizing and adapting BRW sources into quantum optical networks.

\section*{Acknowledgements}
We thank Matthias Covi for assistance with the experimental setup.
This work was supported in part by the ERC, project \textit{EnSeNa} (257531) and by the FWF through project no. I-2065-N27.

\appendix

\section{\label{app1}Modelling the joint spectral properties of signal and idler}
In this appendix we investigate the  joint spectral amplitude (JSA)  of signal and idler and numerically estimate the spectral overlap $\mathcal{O}$, which determines their indistinguishability in the low gain regime.
Following \cite{Grice1997, Keller1997, U'Ren2005} we find that the joint spectral characteristics in a collinear single-pass PDC source are given by
\begin{equation}
f(\omega_s,\omega_i)=\frac{1}{\mathcal{N}}\alpha(\omega_s + \omega_i)\phi(\omega_s,\omega_i) \text{,}
\label{eq:jsca}
\end{equation}
in which $\mathcal{N}$ accounts for the normalization of the JSA via $\int d \omega_{s} d\omega_{i}|f(\omega_s,\omega_i)|^{2} = 1$, $\alpha(\omega_p = \omega_s + \omega_i)$ describes the pump spectrum  in terms of the frequencies $\omega_{\mu}$ ($\mu = p,s,i$) for pump, signal and idler, respectively, and  $\phi(\omega_s,\omega_i)$ is the phasematching (PM) function.
In a Gaussian approximation we can describe the pump amplitude as
\begin{equation}
\alpha(\omega_s + \omega_i) = e^{-\frac{1}{\sigma_p^2}(\omega_s +\omega_i)^{2}}
\label{eq:alpha}
\end{equation}
with $\sigma_p$  being the bandwidth of the pump.  We use a simple PDC model for uniform waveguides \cite{Christ2009, Zhukovsky2012} with constant-valued non-linearity over the whole length of the waveguide and assume that the overlap of spatial modes effectively affects only its strength.
Thus, in the single-pass configuration we can write the PM function as \cite{Grice2001}
\begin{align}
\phi(\omega_s,\omega_i) &= \textnormal{sinc} \bigg(\frac{L}{2} \Delta k(\omega_s,\omega_i) \bigg) e^{i \frac{L}{2}\Delta k(\omega_s,\omega_i)} \nonumber \\
& \approx e^{-\gamma\frac{L^2}{4} \Delta k^2(\omega_s,\omega_i)} e^{i \frac{L}{2} \Delta k(\omega_s,\omega_i)} \text{,}
\label{eq:phi1}
\end{align}
in the final form of which we have used a Gaussian approximation for the sinc-function. In \eref{eq:phi1} $L$ denotes the BRW length, $\Delta k(\omega_s,\omega_i)= k_p(\omega_s + \omega_i) - k_s(\omega_s) - k_i(\omega_i)$ describes the phase mismatch in terms of $k_{\mu}(\omega_{\mu}) = n_{\mu} \omega_{\mu}/c$ with $n_{\mu}$ being the effective refractive index and $c$ the speed of light, while $\gamma \approx 0.193$ adjusts the width of the approximated PM function. Performing a Taylor expansion of $\Delta k$ to the second order  at a phasematched point $\omega_{p}^0 = \omega_{s}^0 + \omega_{i}^0$ we can write the phase mismatch as
\begin{equation}
\Delta k(\omega_s,\omega_i) \approx \kappa_s \nu_s + \kappa_i \nu_i + \Lambda_s \nu_s^2 + \Lambda_i \nu_i^2 - \Lambda_p \nu_s \nu_i,
\label{eq:dk}
\end{equation}
in which the detunings are defined as $\nu_{\mu}=\omega_{\mu} - \omega_{\mu}^0$.  In \eref{eq:dk} $\kappa_{\mu}=k'_{\mu}(\omega_{\mu}^0) - k'_{p}(\omega_{p}^0) = 1/v_{g(\mu)} - 1/v_{g(p)}$ is determined by the group velocity mismatch of the downconverted photons and the pump photon, while $\Lambda_{s,i}=\frac{1}{2}k''_{s,i}(\omega_{s,i}^0)-\frac{1}{2}k''_p(\omega_p^0)$ and $\Lambda_{p}=k''_{p}(\omega_{p}^0)$ are related to the group velocity dispersions.

For our simulation we substitute  \eref{eq:alpha}-\eref{eq:dk} into \eref{eq:jsca} and evaluate the JSA. From a commercially available solver (Mode Solutions \cite{Mode06}) we obtain  for the PDC  process in our BRW (in \fref{fig:exp}(a))  the dispersion properties  listed in \tref{tab:parameters}. 
In \fref{fig:jsa} we show the joint spectral intensity (JSI), $|f(\omega_s,\omega_i)|^{2}$, as a function of the signal and idler frequencies $f_{s,i}=\omega_{s,i}/2\pi$. We evaluated JSI at the extracted degeneracy point of $f_s^0$=$f_i^0$=\SI{193.3}{\tera\hertz} corresponding to \SI{1551.1}{\nano \meter} with a pump having a \SI{0.25}{\nano\meter} broad spectrum. The simulated degeneracy point is very close to the measured one found at \SI{1544.1(8)}{\nano\meter}. We believe this slight discrepancy is due to the experimental limitations in the BRW growth process concerning the accuracy at which the refractive index and the thickness of the layers can be controlled.
\begin{table}[!hbt]
  \centering
  \caption{Parameters of the pump (p), signal (s) and idler (i) mode extracted from numerical simulations \cite{Mode06}.}
    \begin{widetable}{\linewidth}{ccc|cc|ccc}
    \multicolumn{3}{c|}{$v_{g(\mu)}$ \SI{}{(\micro \meter/\pico\second)}} & \multicolumn{2}{c|}{$\kappa_\mu$ \SI{}{(10^{-3}\pico\second/\micro \meter)}} & \multicolumn{3}{c}{$\Lambda_\mu$ \SI{}{(10^{-6}\pico\second^2/\micro \meter)}} \\
    \hline
    p & s & i & s & i & p & s & i \\
    74.0  & 90.1  & 90.4  & -2.40 & -2.44 & 5.74  & -2.16 & -2.17 \\ 
    \hline
    \hline   
    \end{widetable}%
  \label{tab:parameters}%
\end{table}

Due to small difference in the group velocities of the signal and idler photons in the vicinity of the degeneracy point, the tilt of the PM function $\theta \approx \arctan(\kappa_s / \kappa_i)$ deviates from that of perfect anti-correlation by about \SI{0.5}{\degree}. Altogether, our simulated JSI is slightly asymmetric around the degeneracy point, leading to different spectral properties of signal and idler centered at \SI{1567}{\nano \meter} and \SI{1535}{\nano \meter}, respectively. Both marginal spectra are originally approximately \SI{90}{\nano \meter} wide. Thus, they are two orders of magnitude broader than the linewidth of the JSI at its degeneracy point, being about \SI{0.6}{\nano \meter} as shown in the inset in \fref{fig:jsa}. Therefore, even in the case of spectral filtering we expect a highly multimodal PDC emission \cite{Brida2009} from our BRW sample and almost perfectly spectrally indistiguishable twin beams.

\begin{figure}[!hbt]
\centering
\includegraphics[scale=1.1]{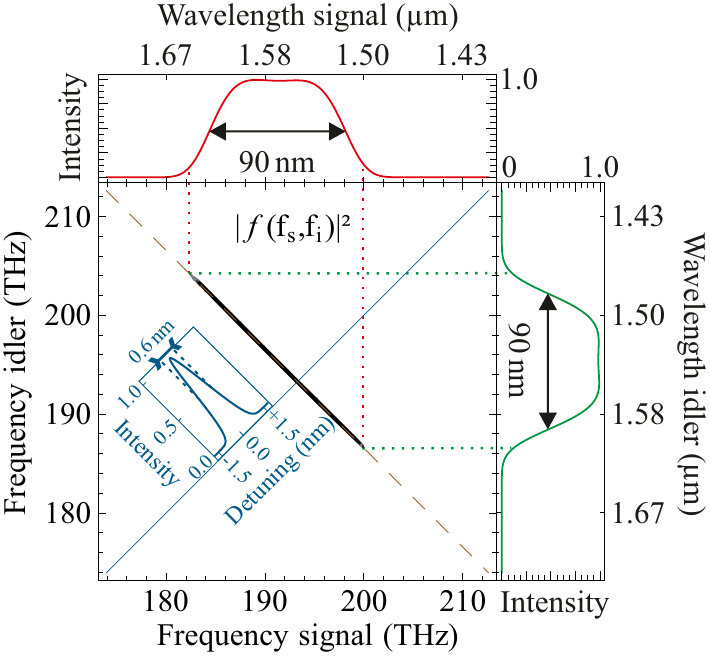}
\caption{(Color online) A contour plot of JSI with respect to the signal and idler frequencies/wavelengths. The inset illustrates the width of the JSI in terms of detuning from the degeneracy point. The red and green solid curve represent the marginal spectra. The diagonal (blue solid line) and the anti-diagonal (brown dashed line) as well as the dotted lines provide guides for the eye.}
\label{fig:jsa}
\end{figure}
 
The spectral overlap is determined by \cite{Avenhaus2009} 
$
\mathcal{O}=  \int \int d\omega_s d\omega_i f(\omega_s,\omega_i)f^*(\omega_i,\omega_s)
$
and depends remarkably on the group velocity mismatch.
As in our case signal and idler travel with  slightly different group velocities, their wavepackets are temporally shifted, which is evident from the phase term of the JSA. For the unfiltered JSA we determine a spectral overlap of only about \SI{26}{\percent}, while \SI{76}{\percent} could be achieved if the temporal mismatch was corrected.  We restrict the influence of the group velocity mismatch by spectral filtering close to the JSA degeneracy point and, therefore, we expect a spectral overlap of  about \SI{98}{\percent} and \SI{83}{\percent} when filtering with a \SI{12}{\nano \meter} (\SI{1.5}{\tera \hertz}) Gaussian and a \SI{40}{\nano \meter} (\SI{5.0}{\tera \hertz}) super-Gaussian BPF, respectively. These values are in good accordance with our experimentally determined results in Sec.~\ref{sec:results2}.

\section{\label{app2}Quantum interference experiment with a twin beam state}
%
We utilize the description of parametric downconversion as multimode  squeezer in order to estimate the effect of the higher photon-number contributions on a quantum interference between signal and idler \cite{Christ2011, Schlawin2013}.
We start by considering the configuration in \fref{fig:scheme}, in which the bunching takes place at a symmetric beam splitter having a transmission of $T=1/2$ corresponding to the case of recording the minimum amount of coincidences in our coalescence measurement. Prior that signal and idler are subjected to losses as is the case also in our experimental realization.  
\begin{figure}[!hbt] \centering
 \includegraphics[width = 0.28\textwidth]{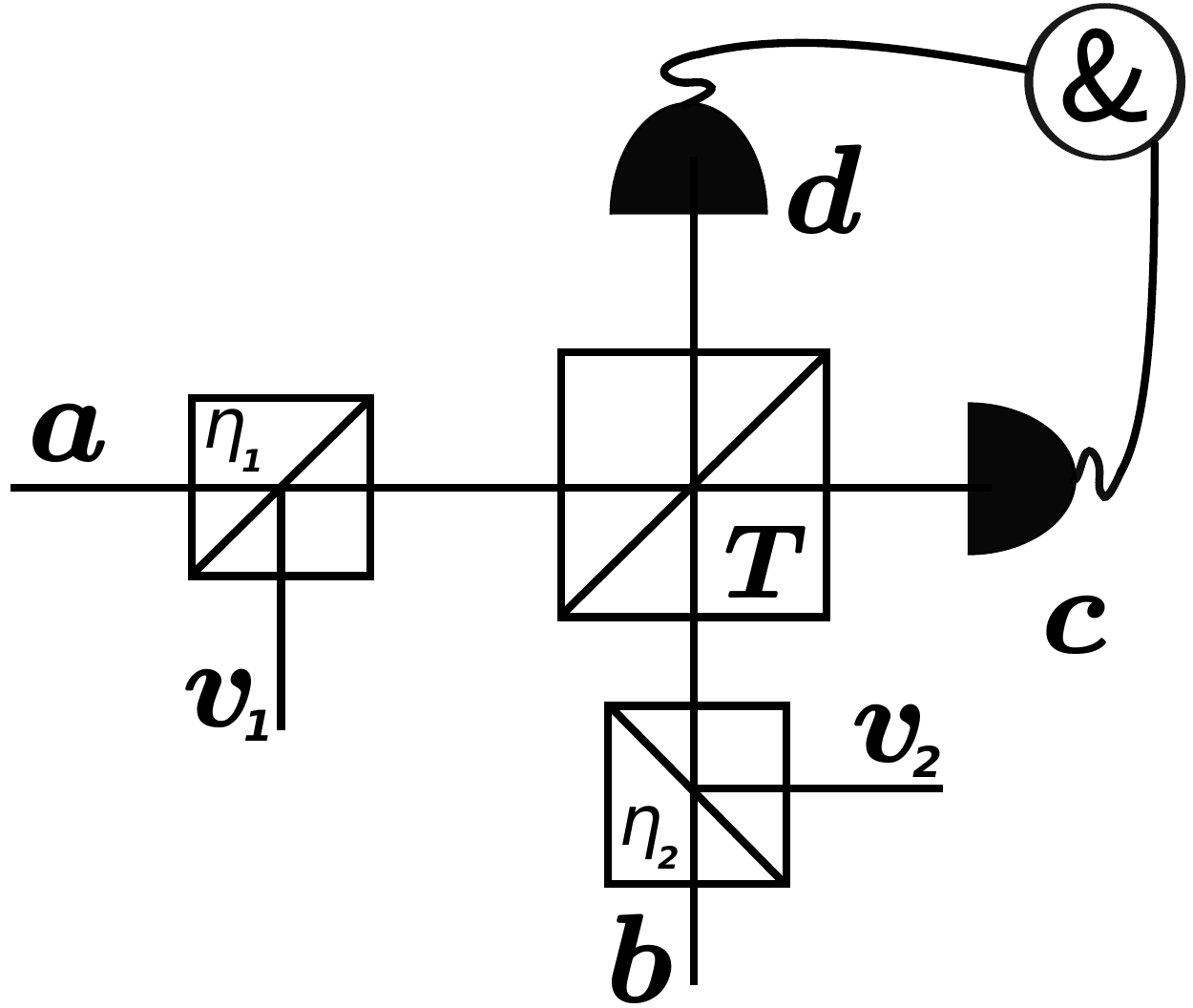}%
 \caption{\label{fig:scheme}Model of bunching experiment with a twin beam state. The twin beams send to the input arms $a$ and $b$ are degraded due to losses at the beam splitters with transmissions $\eta_{1}$ and $\eta_{2}$, to which the optical vacuum modes  $v_{1}$ and $v_{2}$  are coupled, respectively. Bunching occurs at the beam splitter with transmission $T$ and coincidences are counted between the two output arms $c$ and $d$. For more details see text.}
\end{figure}

We define the transformation between the system input arms, $a$ and $b$, and output arms, $c$ and $d$, in time $t$ as
\begin{align}
\label{eq:BS_c}
\op{c}(t_{1}) = 1/\sqrt{2}& \left[ \sqrt{\eta_{1}}\op{a}(t_{1}) + \sqrt{1-\eta_1}\op{v}_{1}(t_{1}) \right . \nonumber \\
                           +&\left.  \sqrt{\eta_2}\op{b}(t_{1}) + \sqrt{1-\eta_2}\op{v}_{2}(t_{1}) \right]  \\
\op{d}(t_{2}) = 1/\sqrt{2}&\left[\sqrt{\eta_{1}} \op{a}(t_{2}) + \sqrt{1-\eta_1}\op{v}_{1}(t_{2})  \right .\nonumber \\
		- &\left. \sqrt{\eta_2}\op{b}(t_{2}) -\sqrt{1-\eta_2}\op{v}_{2}(t_{2}) \right], 
\label{eq:BS_d}
\end{align}
which expresses the photon annihilators $\op{c}(t_{\varsigma})$ and $\op{d}(t_{\varsigma})$ ($\varsigma = 1,2$) at the beam splitter output ports in terms of those of the inputs $\op{a}(t_{\varsigma})$ and $\op{b}(t_{\varsigma})$ and the vacuum modes $\op{v}_{1,2}(t_{\varsigma})$. The detection efficiencies correspond to the transmission coefficients $\eta_{1,2}$.
Further, we assume that our detectors have a spectrally broad response and their detection windows are much longer than the duration of the generated pulsed wavepackets.  Thus, the coincidence rate can be evaluated via \cite{Grice1997}
\begin{equation}
\mathcal{R} \propto \int dt_{1} \int dt_{2}  \ \braket{\opd{c}(t_{1}) \opd{d}(t_{2}) \op{d}(t_{2}) \op{c}(t_{1})},
\label{eq:R}
\end{equation}
the integrand in which describes the probability of a coincidence at times $t_{1}$ and $t_{2}$ between the two detectors.

Our goal is to rewrite \eref{eq:R} in terms of the beam splitter input operators and then evaluate the expectation values regarding the desired input state. By plugging the beam splitter transformations in \eref{eq:BS_c} and \eref{eq:BS_d} together with their conjugates to \eref{eq:R} and utilizing the Fourier transformations given by $\op{a}(t) = \frac{1}{\sqrt{2\pi}}\int d\omega \  \op{a}(\omega) e^{-i\omega t}$ and $\op{b}(t) =  \frac{1}{\sqrt{2\pi}} \int d\omega \  \op{b}(\omega) e^{-i\omega t}$ with $\omega$ being the optical angular frequency, only few terms survive and we can write down the coincidence rate in the form
\begin{align}
\mathcal{R} \propto  \frac{1}{4} \ &\big< \int d\omega \int d\tilde{\omega} \
 \big[ \eta^{2}_{1}\opd{a}(\omega) \opd{a}(\tilde{\omega}) \op{a}(\tilde{\omega})\op{a}(\omega)   \nonumber \\
 - &\eta_{1}\eta_{2} \opd{a}(\omega) \opd{a}(\tilde{\omega}) \op{b}(\tilde{\omega})\op{b}(\omega)   \nonumber \\
 -&2\eta_{1}\eta_{2}  \opd{a}(\omega) \opd{b}(\tilde{\omega}) \op{a}(\tilde{\omega})\op{b}(\omega)  \nonumber \\
 +&2\eta_{1}\eta_{2}  \opd{a}(\omega) \opd{b}(\tilde{\omega}) \op{b}(\tilde{\omega})\op{a}(\omega)  \nonumber \\
 -&\eta_{1}\eta_{2}   \opd{b}(\omega) \opd{b}(\tilde{\omega}) \op{a}(\tilde{\omega})\op{a}(\omega)    \nonumber \\
  + &\eta_{2}^{2}  \opd{b}(\omega) \opd{b}(\tilde{\omega}) \op{b}(\tilde{\omega})\op{b}(\omega)\big]\  \big> 
\label{eq:R_omega}
\end{align}
in which we have carried out the integration over time  by extending the limits to infinity ($\delta(\omega) = 1/(2\pi)\int_{-\infty}^{\infty} dt_{\mu}e^{it_{\mu}\omega}$ ).

Now, we re-express \eref{eq:R_omega} in terms of broadband detection modes that correspond to those of our downconverter. The multimode squeezed state sent to the input arms $a$ and $b$ of the beam splitter is defined via the unitary  squeezing operator $\op{S}_{a,b}$ as \cite{Christ2011}
\begin{align}
\ket{\Psi} =  \op{S}_{a,b}\ket{0}  =e^{\sum_{k} r_{k} \opd{A}_{k}\opd{B}_{k} -h.c.}\ket{0} ,
\end{align}
in which the real valued squeezing strength $r_{k} = \mathcal{B}\lambda_{k}$ is related to the gain of the PDC process $\mathcal{B}$ and to the Schmidt modes $\lambda_{k}$ ($\sum_{k}\lambda^{2}_{k}= 1$) of the joint spectral correlation function of signal and idler given by $f(\omega_{s}, \omega_{i}) = \sum_{k}\lambda_{k} \varphi_{k}(\omega_{s}) \phi_{k}(\omega_{i})$.
With the help of the  two sets of orthonormal basis functions $\{\varphi_{k}\} $ and $\{\phi_{k}\}$ for signal and idler, respectively, we define the k-th mode sent to the input arm $a$ as
\begin{align}
\opd{A}_{k} &= \int d\omega \ \varphi_{k}(\omega)\ \opd{a}(\omega),
\end{align}
for which the following relations hold:
\begin{align}
 \int d\omega \ \varphi^{\star}_{k}(\omega)  \varphi_{k^{\prime}}(\omega)& = \left\{
\begin{array}{rl}
1 & \text{if } k= k^{\prime},\\
0 & \text{otherwise}\\
\end{array} \right. \quad \textrm{and} \nonumber \\
\sum_{k}\varphi_{k}(\omega)\varphi^{\star}_{k}(\tilde{\omega})  &= \delta(\omega-\tilde{\omega}).
\label{eq:conditions}
  \end{align}
Similarly, the k-th mode sent to the input arm $b$ can be written as
\begin{align}
\opd{B}_{k} &= \int d\omega \ \phi_{k}(\omega)\ \opd{b}(\omega),
\end{align}
the basis functions in which obey conditions similar to those in \eref{eq:conditions}.
The broadband mode transformations of the k-th input modes can be presented in the form \cite{Christ2011}
\begin{align}
\label{eq:SAS}
 \opd{S}_{a,b} \op{A}_{k}  \op{S}_{a,b} & = \textrm{cosh}(r_{k}) \op{A}_{k} + \textrm{sinh}(r_{k})\opd{B}_{k} \\
 \opd{S}_{a,b} \op{B}_{k}  \op{S}_{a,b} & = \textrm{cosh}(r_{k}) \op{B}_{k} + \textrm{sinh}(r_{k})\opd{A}_{k}.
 \label{eq:SBS}
\end{align}
In the following we consider only the case of weak squeezing and  approximate   $\textrm{sinh}(r_{k}) \approx r_{k} = \mathcal{B}\lambda_{k}$ and $\textrm{cosh}(r_{k}) \approx 1$. Further, we estimate the mean photon number in the both input arms as $\braket{n} = \sum_{k} \textrm{sinh}^{2}(r_{k}) \approx  \mathcal{B}^{2}$.

In order to transform \eref{eq:R_omega} to the broadband-mode picture, we require the identities
\begin{align}
& \opd{a}(\omega) = \sum_{k} \opd{A}_{k} \varphi^{\star}_{k}(\omega) \quad \textrm{and} \nonumber \\
& \opd{b}(\omega) = \sum_{k} \opd{B}_{k} \phi^{\star}_{k}(\omega).
\end{align}
Thence, we re-express the coincidence rate in \eref{eq:R_omega} as
\begin{align}
&\mathcal{R} \propto  \frac{1}{4}\bigg<\bigg[ \eta^{2}_{1} \sum_{k,n} \opd{A}_{n}\opd{A}_{k}\op{A}_{k}\op{A}_{n} 
 \nonumber \\
&- \eta_{1}\eta_{2}\hspace{-1.5ex} \sum_{k,n,l,m} \hspace{-1.5ex}\opd{A}_{n}\opd{A}_{k}\op{B}_{l}\op{B}_{m}   %
      \hspace{-0.75ex}   \int \hspace{-0.75ex}d\omega \varphi^{\star}_{n}(\omega) \phi_{m}(\omega)%
       \hspace{-0.75ex}   \int \hspace{-0.75ex}d \tilde{\omega} \varphi^{\star}_{k}(\tilde{\omega})  \phi_{l}(\tilde{\omega}) 
          \nonumber \\
&-2\eta_{1}\eta_{2}\hspace{-1.5ex}\sum_{k,n,l,m} \hspace{-1.5ex} \opd{A}_{n}\opd{B}_{k}\op{A}_{l}\op{B}_{m} %
        \hspace{-0.75ex} \int \hspace{-0.75ex}d\omega \varphi^{\star}_{n}(\omega) \phi_{m}(\omega)%
         \hspace{-0.75ex} \int \hspace{-0.75ex}d \tilde{\omega} \phi^{\star}_{k}(\tilde{\omega})  \varphi_{l}(\tilde{\omega}) 
          \nonumber \\
&+2\eta_{1}\eta_{2}\sum_{k,n} \opd{A}_{n}\opd{B}_{k}\op{B}_{k}\op{A}_{n} 
\nonumber \\
&-\eta_{1}\eta_{2}\hspace{-1.5ex}\sum_{k,n,l,m} \hspace{-1.5ex} \opd{B}_{n}\opd{B}_{k}\op{A}_{k}\op{A}_{n}  %
        \hspace{-0.75ex} \int \hspace{-0.75ex}d\omega \phi^{\star}_{n}(\omega) \varphi_{m}(\omega)%
        \hspace{-0.75ex}  \int \hspace{-0.75ex}d \tilde{\omega} \phi^{\star}_{k}(\tilde{\omega})  \varphi_{l}(\tilde{\omega})  
          \nonumber \\
&+ \eta^{2}_{2} \sum_{k,n} \opd{B}_{n}\opd{B}_{k}\op{B}_{k}\op{B}_{n} \bigg]\bigg> . 
\label{eq:R_broadband}
\end{align}
We directly recognize that several terms in \eref{eq:R_broadband} correspond to Glauber correlation functions $\mathcal{G}(w,\upsilon) = \braket{:(\sum_{q}\opd{A}_{q}\op{A}_{q})^{w}(\sum_{q^{\prime}}\opd{B}_{q^{\prime}}\op{B}_{q^{\prime}})^{\upsilon}:}$ with indices $w$ and $\upsilon$ describing the order of the correlation for the twin beam modes $a$ and $b$, respectively \cite{Christ2011}. Thence, when disregarding the losses, the expectation values in the first and last terms deliver $\mathcal{G}(2,0) =  \mathcal{G}(0,2) = \braket{n}^{2}[1+\frac{1}{K}]$, and in the fourth term  $ \mathcal{G}(1,1) = \braket{n}^{2}[1+\frac{1}{K}]+\braket{n}$, where $K$ corresponds to the effective number of excited modes ($K = 1/\sum_{k}\lambda^{4}_{k}$) \cite{Christ2011}. The rest of the mean values can be evaluated by plugging in the transformations from \eref{eq:SAS} and \eref{eq:SBS} together with their hermitian conjugates. While the second and fifth terms vanish, the third term delivers 
\begin{align}
&\big< \hspace{-0.75ex} \sum_{k,n,l,m} \hspace{-0.5ex} \opd{A}_{n}\opd{B}_{k}\op{A}_{l}\op{B}_{m} %
        \int \hspace{-0.5ex}d\omega \varphi^{\star}_{n}(\omega) \phi_{m}(\omega)%
         \int \hspace{-0.5ex}d \tilde{\omega} \phi^{\star}_{k}(\tilde{\omega})  \varphi_{l}(\tilde{\omega}) \hspace{0.25ex} \big> \nonumber \\
 & =  \braket{n} \mathcal{O}  + \braket{n}^{2} \mathcal{A},
\end{align}
in which 
\begin{align}
\mathcal{O} &= \int d \omega \int d \tilde{\omega} \ f^{\star}(\omega, \tilde{\omega})f(\tilde{\omega}, \omega)  \\
&= \sum_{k,n}  \lambda_{n}\lambda_{k}  \int \hspace{-0.5ex}d\omega \varphi^{\star}_{n}(\omega) \phi_{k}(\omega) \int \hspace{-0.5ex}d \tilde{\omega} \phi^{\star}_{n}(\tilde{\omega})  \varphi_{k}(\tilde{\omega})\nonumber
\end{align} 
describes the spectral overlap between signal and idler  and
\begin{align}
\mathcal{A} &=  \int d \omega \int d \tilde{\omega} \ g_{s}(\omega, \tilde{\omega})g_{i}(\tilde{\omega}, \omega)  \\
&= \sum_{k,n}  \lambda^{2}_{n}\lambda^{2}_{k}  \int \hspace{-0.5ex}d\omega \varphi^{\star}_{n}(\omega) 
  \phi_{k}(\omega) \int \hspace{-0.5ex}d \tilde{\omega} \phi^{\star}_{k}(\tilde{\omega})  \varphi_{n}(\tilde{\omega})\nonumber
  \end{align}
determines the overlap of signal and idler spectral densities that are given by
\begin{align}
&g_{s}(\omega, \tilde{\omega}) = \int d \omega_{i} \ f^{\star}(\omega, \omega_{i})f(\tilde{\omega}, \omega_{i}) \quad \textrm{and} \\
&g_{i}(\tilde{\omega}, \omega) = \int d \omega_{s} \ f^{\star}(\omega_{s}, \tilde{\omega})f(\omega_{s}, \omega).
\end{align}
We note that if the spectral densities of signal and idler  are equal this term will end up giving the purity of the photon wavepacket $1/K$.

Finally, we determine an expression for the visibility of our quantum interference experiment in Sec.~\ref{sec:results2}. When  signal and idler are expected to bunch, we can estimate the rate of the coincidences according to  \eref{eq:R_broadband} as
\begin{align}
\mathcal{R}_{\textrm{min.}} \propto &\frac{1}{4} \braket{n}^{2}(1+\frac{1}{K})(\eta_{1}^{2}+ \eta_{2}^{2}) \nonumber\\
+&\frac{1}{2}\Big(\braket{n} + \braket{n}^{2}(1+\frac{1}{K})\Big)\eta_{1}\eta_{2}  \nonumber\\
-&\frac{1}{2}\Big(\braket{n}\mathcal{O} + \braket{n}^{2}\mathcal{A} \Big)\eta_{1}\eta_{2}.
\end{align}
This rate is compared with the one obtained when the signal and idler beams are separated deterministically. By using the same model as above but assuming a beam splitter with $T = 1$ in \fref{fig:scheme}, we gain
\begin{equation}
\mathcal{R}_{\textrm{max.}} \propto \bigg(\hspace{-0.75ex}\braket{n} + \braket{n}^{2}\Big(1+\frac{1}{K}\Big) \bigg) \eta_{1}\eta_{2}.
\end{equation}
The visibility is then given by
\begin{align}
\label{eq:visibility}
\mathcal{V} &= \frac{\mathcal{R}_{\textrm{max.}} - \mathcal{R}_{\textrm{min.}}}{\mathcal{R}_{\textrm{max.}}+\mathcal{R}_{\textrm{min.}}}  \nonumber \\
&\approx \frac{[1+\mathcal{O}]+\braket{n} \Big(1-\frac{1}{2}(\frac{\eta_{1}}{\eta_{2}}+\frac{\eta_{2}}{\eta_{1}})\Big)}{[3-\mathcal{O}]+ 3\braket{n} + \frac{1}{2}\braket{n} (\frac{\eta_{1}}{\eta_{2}}+\frac{\eta_{2}}{\eta_{1}}) },
\end{align}
in the final form of which we have used the approximation that our PDC process is highly multimodal. Moreover, we see from \eref{eq:visibility} that if the twin beams are detected with largely different efficiencies, this causes an unbalance due to which the visibility is degraded. In our case the Klyshko efficiencies, with which signal and idler are detected, are close to each other and in good approximation we can model the measured visibilities in terms of the mean photon number as
\begin{equation}
\mathcal{V} \approx\frac{1+\mathcal{O}}{3-\mathcal{O}+ 4\braket{n}}
\end{equation}
that contains in addition to the spectral overlap known for a true photon-pair state \cite{Avenhaus2009} a degradation in the visibility due to the higher photon-number contributions.

\section*{References}
\bibliographystyle{unsrt} 
\bibliography{brw} 

\end{document}